\documentclass[traditabstract]{aa} 
\usepackage{graphicx}
\usepackage{txfonts}
\begin{document}

   \title{An investigation of magnetic field distortions
          \\ in accretion discs around neutron stars}
   \subtitle{I. Analysis of the poloidal field component}

   \author{L. Naso \inst{1}\fnmsep\inst{2}
          \and
          J.C. Miller\inst{1}\fnmsep\inst{3}
          }
   \institute{SISSA and INFN, via Bonomea 265, 34136 Trieste, Italy
         \and
             Department of Physics, University of Padova, via Marzolo 8, 35131 Padova, Italy
         \and                         
             Department of Physics (Astrophysics), University of Oxford, Keble
Road, Oxford OX1 3RH, UK
          }

   \date{Received 14 December 2009 / Accepted 07 June 2010}

 
  \abstract {We report results from calculations investigating stationary 
 magnetic field configurations in accretion discs around magnetised neutron 
 stars. Our strategy is to start with a very simple model and then 
 progressively improve it, providing complementary insight into results 
 obtained with large numerical simulations. In our first model, presented 
 here, we work in the kinematic approximation and consider the stellar magnetic
field as being a dipole aligned with the stellar rotation axis and 
 perpendicular to the disc plane, while the flow in the disc is taken to be 
 steady and axisymmetric. The behaviour in the radial direction is then 
 independent of that in the azimuthal direction. We investigate the distortion 
 of the field caused by interaction with the disc matter, 
 solving the induction equation numerically in full 2D. The influence of
turbulent diffusivity and fluid velocity on the poloidal field configuration is
analysed, including discussion of outflows from the top and bottom of the 
 disc. We find that the distortions increase with increasing magnetic Reynolds 
 number $R_{{\rm m}}$ (calculated using the radial velocity). However, a 
 single global parameter does not give an adequate description in different 
 parts of the disc and we use instead a `magnetic distortion function' 
 $D_{{\rm m}}(r,\theta)$ (a magnetic Reynolds number defined locally). Where 
 $D_{{\rm m}}\ll1$ (near to the inner edge of the disc) there is little 
 distortion, but where $D_{{\rm m}} > 1$ (most of the rest of the disc), there
is considerable distortion and the field becomes weaker than the dipole would
have been. Between these two regions, there is a transition zone where the field is amplified and can have a local minimum and maximum. The location of this zone depends sensitively on the diffusivity. The results depend very little on the boundary conditions at the top of the disc.}

   \maketitle
%

\section{Introduction}\label{sec:INTRO}

 With this paper, we begin a study of the properties of accretion discs around 
 magnetised neutron stars. We are interested in two kinds of system: X-ray 
 pulsars and old neutron stars in the process of being spun-up (recycled) to 
 become millisecond pulsars. Here we focus mostly on these recycled 
 pulsars, for which the distortions of the magnetic field are larger and occur 
 at smaller distances from the central object, making the effects easier to 
 see.

 The study of accretion onto magnetised neutron stars was pioneered by Ghosh and Lamb who, in a series of 3 papers (Ghosh et al. \cite{GL77}; Ghosh \& Lamb \cite{GL79a}, hereafter referred to as GL; Ghosh \& Lamb \cite{GL79b}), investigated the flow of accreting matter and the magnetic field configuration. They were able to estimate the total torque exerted on the neutron star by the accretion disc but, for doing this, they needed to make a number of quite drastic assumptions. In particular they used a dipolar profile for the poloidal component of the magnetic field and an ad hoc prescription for the azimuthal component, which was later shown to be inconsistent with having a stable disc beyond the corotation point (Wang \cite{W87}).

 Wang (\cite{W87}) and Campbell (\cite{C87}, \cite{C92}) obtained an analytic 
expression for the azimuthal component of the magnetic field, making the 
assumption that the poloidal component is dipolar. They solved the induction 
equation to find a steady solution for an axisymmetric magnetic field and 
found that the generation of toroidal field is due only to the vertical 
gradient of the azimuthal velocity $v_\phi$. By further assuming that the disc 
is Keplerian\footnote{Campbell also considered non-Keplerian flow in the inner
part of the disc.} ($\Omega=\Omega_{\rm K}$) and that the magnetosphere is
corotating with the star (with angular velocity $\Omega_{\rm s}$), they then
obtained $B_\phi\propto (\Omega_{\rm K}-\Omega_{\rm s}) \, B_z$.

 Miller \& Stone (\cite{MS97}) performed some numerical simulations with a 2D 
model, including the effects of non-zero resistivity. Their simulations 
confirmed that there is not a total screening of the magnetic field from the 
accretion disc (as had initially been assumed by Ghosh et al. \cite{GL77}) and also that the poloidal component can be very different from a dipolar field. Agapitou \& Papaloizou (\cite{AP00}) also found that the poloidal component of the field can differ significantly from the dipole field of the central star, and that the magnetic torque can be much smaller than that estimated assuming $B_z\sim B_{{\rm dip}}$.

 Our aim here is to improve on these analyses and to find a stationary 
configuration for the magnetic field inside the disc and in the surrounding 
corona, without making any leading order expansion or vertical integration in 
the induction equation, without making any assumptions about the poloidal 
component of the magnetic field inside the disc and by using a more general 
profile of the velocity field (with all of the components being allowed to be 
non-zero).

 Many investigators have simulated the magnetosphere--disc interaction, 
solving the full set of the MHD equations (see e.g. Romanova et al. 
\cite{Rom02}; Kulkarni \& Romanova \cite{KR08}). The work that we are 
presenting here should not be seen as being in competition with these 
analyses, but rather as being complementary to them, by providing both a 
useful test case and a means to better understand the underlying processes. 
Our strategy is to use a succession of simplified models (of which this paper 
presents the initial one) becoming progressively more sophisticated, and to 
proceed step by step so as to fully understand the effect of each successive 
additional feature as it is introduced. In large-scale numerical works, one 
sees the results of an interacting set of inputs within the scope of the 
adopted model assumptions and numerical techniques. Deconstructing this, so as 
to have a clear conceptual understanding of the role of each of the different 
components, remains a valuable thing to do and an approach which needs to be 
carried on alongside the large-scale simulations. The conceptual papers from 
the 70s and 80s mentioned above, continue to be widely quoted and used as the 
basis for new research (see e.g. Kluzniak \& Rappaport \cite{KR07}) and our 
work here stands in the tradition of refining these approaches.

 The structure of this paper is as follows. After the present 
Introduction, the details of the model are given in Sect. \ref{sec:MOD}, 
and the equations are presented in Sect. \ref{sec:EQ}. As we will show 
there, it is possible to separate the calculation for the poloidal 
component of the magnetic field from that for the toroidal component. In 
this paper we present results for the poloidal field; the toroidal one 
will be discussed in a subsequent paper. The numerical code used is 
described briefly in Sect. \ref{sec:EQ}, and then in more detail in the 
Appendix (only in the online version of the paper). In Sect. \ref{sec:RES} we
present our results and Sect. \ref{sec:CON} contains conclusions.

\section{Model}\label{sec:MOD}

 In this study we consider disc accretion by a neutron star having a dipolar 
magnetic field. We assume that the star is rotating around its magnetic axis, 
and that this axis is perpendicular to the disc plane; also, we assume that 
the fluid flow is steady and has axial symmetry everywhere. We use the 
kinematic approximation and do not consider any dynamo action. Turbulent 
diffusivity is included and the velocity field is not constrained to be 
purely rotational but it is allowed to be non-zero also along the other 
directions. We use spherical coordinates ($r$, $\theta$, $\phi$), with the 
origin being at the centre of the neutron star.

 We assume that at large radii the magnetic pressure is negligible with 
respect to the gas pressure and that the disc can be described there as a 
standard $\alpha$-disc. As one moves inwards, the magnetic field becomes 
progressively stronger and, for the cases considered here, eventually 
dominates over the gas pressure. It is convenient to divide the disc into 
two regions: an outer zone, where the magnetic field does not greatly 
influence the flow, and an inner zone, where the magnetic stresses are 
dominant over the viscous ones and matter begins to leave the disc 
vertically. For a sufficiently strong magnetic field, most of the matter 
eventually leaves the disc following magnetic field lines and the disc is 
disrupted. We note that some equatorial accretion continues to be possible 
even with a rather high stellar magnetic field, as shown by 
Miller \& Stone (\cite{MS97}).

 The precise location of the inner edge of the disc, $r_{\rm in}$, is 
still an open issue: several prescriptions have been suggested, but none 
of them differs very much from the Alfven radius calculated using a 
dipolar magnetic field. For our present model we take this as giving the 
inner edge of the disc; for subsequent models, a possible improvement will 
be to calculate the Alfven radius using the stationary magnetic field 
obtained in this work instead of the dipolar one.

 We note here that, if we take the innermost part of the disc to have 
very high diffusivity, we naturally obtain the same structure as that of the 
GL model, i.e. a narrow inner region followed by a broad outer one. However 
the behaviour of the field in these regions is quite different from that in 
the GL model, as will be shown in the results section.

 We suppose that all around the pulsar there is vacuum, except for where 
we have the disc and the corona, and that the field remains dipolar 
everywhere in this vacuum. In reality, this is not completely correct,
both because the density in the magnetosphere is not zero and because 
between the star and the disc there is the matter which is accreting onto 
the neutron star. For the former we are introducing a low density corona
in order to have a transition zone between the disc and the vacuum. We also
allow the velocity and the diffusivity to have different values in the corona
from those in the disc. For the latter, we suppose that the matter flow is 
perfectly aligned with field lines and that it has very low density, so 
that we can neglect its influence on the magnetic field structure. We are aware that in real astrophysical systems, imposing the dipole condition at the
boundaries is rather drastic since the field is distorted well before reaching
the disc. However at our level of analysis, the results are not very sensitively dependent on this, and we are here focusing on studying the influence on the magnetic configuration of the velocity field and diffusivity. That this is reasonable is confirmed by the fact that the magnetic configuration which we obtain is rather similar to that given by the numerical simulations of other authors (Miller \& Stone 1997), which had a different treatment of the boundary conditions. We will comment more on this in Sect. 
\ref{sec:RES}.

\begin{figure}
\begin{center}
\includegraphics[width=0.35\textwidth]{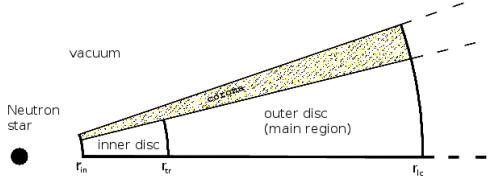}
\end{center}
 \caption{Schematic representation of our model (not to scale). We use 
$r_{\rm in}=10\,r_{{\rm g}}$, $r_{\rm tr}\sim22\,r_{{\rm g}}$ and $r_{\rm 
lc}\sim115\,r_{{\rm g}}$. The opening angles are $8^{\circ}$ for the disc 
alone and $10^{\circ}$ for disc plus corona. The outer disc extends much
further out than the main region shown here: the grid continues until $r_{\rm
out}=750\,r_{{\rm g}}$.}
 \label{fig:model}
\end{figure}

 Summarizing, we divide the surroundings of the central object into four parts (see Fig. \ref{fig:model}): (1) the inner disc, extending from $r_{\rm in}$ out to a transition radius $r_{\rm tr}$ (where the diffusivity changes); (2) the outer disc, extending from $r_{\rm tr}$ to an outer radius $r_{\rm out}$; (3) a corona, above and below the disc; and (4) everything else, which we take here to be vacuum. Within the outer disc, we focus on the {\em main region,} extending from $r_{\rm tr}$ out to the light cylinder at $r_{\rm lc}$; the outer edge of our grid, $r_{\rm out}$, is very much further away than this, so that conditions imposed there do not affect the solution in the region of interest.

 We take the ratio $h/r$ to be constant (where $h$ is the semi-thickness of the disc), and so the entire upper surface of the disc is located at a single value of $\theta$ (as also is the case for the corona). We take an opening angle of $8^\circ$ for the disc (measuring from the equatorial plane to the top of the disc), and $10^\circ$ for the disc plus corona. The disc then has $h/r\sim0.14$.

 Once the magnetic diffusivity $\eta$ and the poloidal velocity $\mathbf{v_{\rm
p}}$ are specified, the poloidal component of the induction equation can be
solved to obtain the configuration of the poloidal field without entering into
details of the toroidal component. As regards the turbulent magnetic diffusivity, we take this to have a constant value $\eta_0$ in the bulk of the
outer disc, and a higher value $\eta_{\rm c} = 100 \, \eta_0$ in the corona,
the inner disc and near to the outer boundary (we join the different parts
smoothly, using error functions). A similar approach was used by Rekowski et al. (\cite{R00}). We take higher values in regions (1) and (3) because the density is lower there and so we expect the effects of turbulence to be enhanced. Moreover in both regions we expect the field not to be influenced very much by the plasma flow and having a larger diffusivity is a way to reduce this influence. However as we move away from the corona into the vacuum, the density drops to zero and turbulence eventually disappears\footnote{It is only the turbulent diffusivity $\eta_{\rm {\mbox{\tiny T}}}$ which disappears, the Ohmic one $\eta_{\rm \mbox{\tiny Ohm}}$ will always be present, therefore in vacuum $\eta=\eta_{\rm \mbox{\tiny Ohm}}\ll\eta_{\rm {\mbox{\tiny T}}}$.}.

 As regards the radial component of the velocity, $v_r$, we use the expression
given for the middle region of $\alpha$-discs \footnote{ For the parameter
values which we are using, both $r_{\rm in}$ and $r_{\rm lc}$ lie within the
``middle region'' of the $\alpha$-disc model.} by Shakura \& Sunyaev (1973). In
Sect. \ref{subs:AN_C}, we find that whenever a dipole field is a solution of
the induction equation, a precise relation must hold between $v_r$ and 
$v_\theta$. We use this relation to calculate $v_\theta$ in the corona, so as to be consistent with the dipole boundary conditions, and put $v_\theta$ to zero inside the disc.

\section{Equations}\label{sec:EQ}
 If one considers mean fields, the induction equation can be written as:
 \begin{equation}
\partial_t\mathbf{B}=\nabla\times\left(  \mathbf{v}\times\mathbf{B} + 
\mathbf{\mathcal{E}} - \eta_{\rm \mbox{\tiny Ohm}}\nabla\times\mathbf{B}\right)
\end{equation}
 where $\eta_{\rm \mbox{\tiny Ohm}}=c^2/4\pi\sigma$ is the Ohmic 
 diffusivity and $\mathbf{\mathcal{E}}$ is the turbulent electromotive 
 force. A common procedure is to expand $\mathbf{\mathcal{E}}$ in terms of 
 the mean field and its derivatives and within the first order smoothing 
 approximation one has $\mathbf{\mathcal{E}} = \alpha \mathbf{B}-\eta_{\rm 
 {\mbox{\tiny T}}}\nabla\times\mathbf{B}$, where the $\alpha \mathbf{B}$ 
 term generates the so-called $\alpha$-effect. In the present paper we 
 neglect this effect, which could however be included in a subsequent 
 dynamo model. This is in line with our approach, which is aiming at 
 understanding, one at a time, the effects of the various elements which 
 characterize the system of an accretion disc around a magnetised neutron 
 star.

The induction equation then reduces to:
\begin{equation}
\partial_t\mathbf{B}=\nabla\times\left(  \mathbf{v}\times\mathbf{B}  - 
\eta\nabla\times\mathbf{B}\right)
\label{eq:IND}
\end{equation}
 where $\eta=\eta_{\rm \mbox{\tiny Ohm}}+\eta_{\rm {\mbox{\tiny 
 T}}}\sim\eta_{\rm {\mbox{\tiny T}}}$, because the turbulent diffusivity 
 is much larger than the Ohmic one.

 After imposing stationarity (i.e $\partial_t[\dots]=0$) and 
axisymmetry (i.e $\partial_\phi[\dots]=0$), the three components of 
Eq. (\ref{eq:IND}) become:
 \begin{eqnarray}
0 &=& \partial_\theta \left\{ \sin\theta \left[ v_r B_\theta - v_\theta 
B_r - \frac{\eta}{r}[\partial_r(rB_\theta) - \partial_\theta B_r]\right] 
\right\} \label{ind_x}\\
0 &=& \partial_r \left\{ r \hspace{0.6cm}\left[ v_r B_\theta - v_\theta 
B_r - \frac{\eta}{r}[\partial_r(rB_\theta) - \partial_\theta B_r] \right] 
\right\}\label{ind_y}\\
0 &=& \partial_r \left\{ r \left[ v_\phi B_r - v_rB_\phi + \frac{\eta}{r} 
\partial_r(rB_\phi) \right] \right\} -\\
\nonumber
&&\partial_\theta \left\{ v_\theta B_\phi - v_\phi B_\theta  -  
\frac{\eta}{r\sin\theta}\partial_\theta (B_\phi \sin\theta) \right\}  
\label{ind_z}
\end{eqnarray}
 The first two of these equations have the same expression inside the 
 large square brackets and we call this $\mathcal{F}$. We can then reduce 
 these two equations to a single equation and write the entire system as:
 \begin{eqnarray}
\label{eq:F}
\mathcal{F}&=&\frac{k}{r\,\sin\theta}\\
\label{eq:Bphi}
0 &=& \partial_r \left\{ r \left[ v_\phi B_r - v_rB_\phi + \frac{\eta}{r} 
\partial_r(rB_\phi) \right] \right\} -\\
\nonumber
&&\partial_\theta\left\{ v_\theta B_\phi - v_\phi B_\theta  -  
\frac{\eta}{r\sin\theta}\partial_\theta (B_\phi \sin\theta) \right\}
\end{eqnarray}
where $k$ is a generic constant.

 We notice that $\mathcal{F}$ contains only the turbulent magnetic 
diffusivity $\eta$, the poloidal components of the velocity field and the 
$r$ and $\theta$ components of the magnetic field. It is clear therefore 
that Eq. (\ref{eq:F}) on its own (plus Maxwell's equation 
$\nabla\cdot\mathbf{B}=0 $) governs the poloidal part of the magnetic 
field and is independent of any azimuthal quantity, while to obtain the 
toroidal component of the magnetic field one has to solve the further 
equation (\ref{eq:Bphi}).

 In this paper we concentrate only on solving for the poloidal field, 
using Eq. (\ref{eq:F}), for which no knowledge of behaviour in the 
$\phi$ direction is required. In a forthcoming paper we will use the 
results obtained here to solve Eq. (\ref{eq:Bphi}) and calculate the 
toroidal field component.

 In order to guarantee that $\nabla\cdot\mathbf{B}=0$ is satisfied and to 
be able to calculate the magnetic field lines easily, we write the 
magnetic field components in terms of the magnetic stream function 
$\mathcal{S}$, which is implicitly defined by:
 \begin{equation}
\label{eq:S_rt}
B_r = \frac{1}{r^2\sin\theta}\,\partial_\theta \, \mathcal{S}(r,\theta)
\hspace{1cm} 
B_\theta = -\frac{1}{r\sin\theta}\,\partial_r \, \mathcal{S}(r,\theta).
\end{equation}
 With this definition, the axisymmetric field $\mathbf{B}$ is always 
 solenoidal and isolines of $\mathcal{S}$ represent magnetic field lines. 
 Substituting these expressions into Eq. (\ref{eq:F}), we obtain an 
 elliptic partial differential equation for $\mathcal{S}$ in terms of $r$ 
 and $\theta$:
 \begin{eqnarray}
\label{eq:pde1}
&& \partial_r^2\mathcal{S} + \frac{1}{r^2}\partial_\theta^2 
\mathcal{S}-\left( \frac{\cot\theta}{r^2} + \frac{v_\theta}{r\,\eta} 
\right)\partial_\theta \mathcal{S} - \frac{v_r}{\eta}\partial_r 
\mathcal{S} = \frac{k}{\eta}
\end{eqnarray}
 where $k$ is the constant introduced in Eq. (\ref{eq:F}) and $v_r$, $v_\theta$ and $\eta$ are non-constant coefficients. This equation can be solved once boundary conditions and values for the coefficients have been specified. As described in Sect. \ref{sec:MOD} we are interested in a configuration where the magnetic field is a pure dipole at the boundaries. This implies $k=0$ and $S=\mu\sin^2\theta/r$, where $\mu$ is the magnetic dipole moment.

\subsection{Velocity field and turbulent diffusivity}\label{subs:VETA}
 
 In Sect. \ref{sec:MOD}, we justified and qualitatively described
the profiles of velocity and diffusivity that we are using. Here we give 
the precise expressions.

For the velocity, we use the expression given by Shakura \& Sunyaev
(\cite{SS73}):
 \begin{equation}
\label{eq:vr}
v_r(r)=2 \times 10^6 \alpha^{4/5} \dot{m}^{2/5} m^{-1/5} (3/r)^{2/5} \left[ 1-(3/r)^{0.5} \right]^{-3/5} \rm{cm}\,\rm{s}^{-1}
\end{equation}
 where the radius $r$ is expressed in units of the Schwarzschild radius, 
$m$ is given in solar mass units, $\dot{m}$ is in units of the critical 
Eddington rate and $\alpha$ is the standard Shakura-Sunyaev viscosity 
coefficient. Using typical values ($\alpha=0.1$, $\dot{m}=0.03$ and 
$m=1.4$) one obtains:

\begin{equation}
 v_r(r)\approx7.3 \times 10^4 (3/r)^{2/5} \left[ 1-(3/r)^{0.5} \right]^{-3/5}
\rm{cm}\,\rm{s}^{-1}
\end{equation}
 For the other component of the poloidal velocity, $v_\theta$, we set this 
to zero in the disc and use a non-zero profile in the corona. The formula 
for this is calculated in Sect. \ref{subs:AN_C}; we anticipate here 
the result:

\begin{eqnarray}
v_\theta = \left\{ \begin{array}{ll}
0 & \textrm{ \hspace{.25cm} in the disc}\\
\frac{1}{2}\,v_r\,\tan\theta & \textrm{ \hspace{.25cm} in the corona}
\end{array} \right.
\end{eqnarray}

 For the diffusivity, as mentioned in Sect. \ref{sec:MOD}, we are 
using a value $\eta=\eta_0$ in the bulk of the outer disc, and a higher 
value, $\eta = \eta_c = 100 \, \eta_0$, in the corona, the inner disc and 
near to the outer boundary. We use combinations of error functions for 
giving a smooth join between the different regions:

\begin{equation}
 \label{eq:eta}
  \eta(r,\theta) = \eta_0 \cdot \left[1 + \frac{1}{2}\cdot(\eta_\theta(\theta)
+ 
  \eta_r(r)) \cdot \left( \frac{\eta_{\rm c}}{\eta_0} - 1 \right) \right]
\end{equation}
where
\begin{eqnarray}
 \label{eq:eta_t}
  \eta_\theta(\theta) &=& \left[1 + \rm{erf}\left(\frac{-\theta+\theta_{c} 
  }{d_\theta} \right) \right]\\
 \label{eq:eta_r}
  \eta_r(r) &=&\left[1 + \rm{erf}\left(\frac{-r+r_{\eta\,{\rm in}} }{d_r}
\right)
  \right] + \left[1 + \rm{erf}\left( \frac{r-r_{\eta\,{\rm out}}}{d_r} \right)
  \right]
\end{eqnarray}
 where $\theta=\theta_{c}$ on the upper surface of the disc, 
$r_{\eta\,{\rm in}} = r_{\rm tr}$ is the radius of the boundary between 
the inner disc and the main region, and $r_{\eta\,{\rm out}}$ is located 
shortly before $r_{{\rm out}}$, far away from the zone of interest. The 
widths of the error functions used for the angular and radial profiles are 
given by $d_\theta$ and $d_r$ respectively. We have used $d_\theta=10^{-3}$ and $d_r=2\,r_{\rm g}$.

\subsection{Dipolar solution, an analytic constraint}\label{subs:AN_C}

In order to obtain a profile for $v_\theta$ we consider the situation 
when, from the top surface of the corona (at $\theta=\theta_{{\rm surf}}$) 
down to some $\theta = \tilde{\theta}$, the stationary magnetic field is 
dipolar, i.e. $(B_r,B_\theta,B_\phi) = \left( 
\frac{2\mu\,\cos\theta}{r^3}, \frac{\mu\,\sin\theta}{r^3},0\right)$ where 
$\mu$ is the magnetic dipole moment. Then Eq. (\ref{eq:IND}) 
becomes:
 \begin{equation}
\label{eq:ind_dip}
\nabla\times\left(  \mathbf{v}\times\mathbf{B} \right) = 0
\end{equation}
 Following a procedure similar to that used for obtaining Eqs. (\ref{eq:F}) and (\ref{eq:Bphi}), we go to spherical coordinates, write out the three component equations and group the poloidal terms. This gives:
 \begin{eqnarray}
\label{eq:an_pol}
v_r\, B_\theta - v_\theta\, B_r = \frac{\tilde{k}}{r\,\sin\theta} 
\label{vel_pol} \\
\label{eq:an_tor}
\partial_r ( r\,v_\phi\,B_r) =  -\partial_\theta ( v_\phi B_\theta ) 
\label{vel_tor}
\end{eqnarray}
 where $\tilde{k}$ is a generic constant. In order to calculate it we 
consider a path with constant $\theta$, e.g. $\theta = \theta^* \in 
[\theta_{\rm{surf}}, \tilde{\theta}]$; along this path equation 
(\ref{eq:an_pol}) gives:
 \begin{eqnarray}
\label{eq:v_tmp}
v_r\,\left( \frac{\mu\,\sin\theta^*}{2} \right) - v_\theta \, \left( 
\mu\,\cos\theta^* \right) &=& r^2\,\left( 
\frac{\tilde{k}}{\sin\theta^*}\right)
\end{eqnarray}
 where all of the terms in the brackets are constant. We do not know the 
exact profiles of $v_r$ and $v_\theta$, but it is not plausible that the 
left hand side increases as $r^2$ and so we need to choose $\tilde{k}=0$. 
Therefore from the last equation we get:
 \begin{eqnarray}
v_r\,\frac{\sin\theta}{2} = v_\theta\, \cos\theta \rightarrow v_\theta = 
\frac{1}{2}\,\tan\theta\,v_r
\label{eq:vrvt}
\end{eqnarray}
 Equation (\ref{eq:vrvt}) implies not only that if there is a non-zero 
radial velocity then there must be a non-zero vertical velocity as well, 
but also that the vertical velocity is larger than the radial one 
calculated at the same location (for $\theta=81~^\circ $ one has $v_\theta 
\sim 13\,v_r$).

 The behaviour of the azimuthal velocity $v_\phi$ can be investigated 
using Eq. (\ref{eq:an_tor}). As expected $v_\phi=0$ is a possible 
solution but $v_\phi=v_{\rm{Kep}}$ is not, meaning that having a dipolar 
field is not consistent with having a Keplerian angular velocity profile, 
whereas it is consistent with having no rotation at all. It is 
interesting that there are also some non trivial profiles which are 
solutions. For example $v_\phi \propto r^{-\delta/2}\cdot 
\sin^{3+\delta}\theta$, which is equivalent to $\Omega \propto
r^{-\alpha/2} \sin^\alpha\theta$, gives a set of possible solutions. Therefore if one supposes that $v_\phi$ decreases with $r$ as a power law, then it must have also a dependence on $\theta$ in order for the magnetic field to be dipolar. We recall that in the works of Campbell (\cite{C87}, \cite{C92}) and Wang (\cite{W87}) it is precisely the vertical gradient of the angular velocity that produces the toroidal field. Here we have shown that it is possible to have a non-zero vertical gradient of the angular velocity and still have a zero toroidal magnetic field. However, we stress that we are not giving physical explanations for having these kinds of velocity profiles. We will comment further on the profile of the angular velocity in the forthcoming paper where we will solve Eq. (\ref{eq:Bphi}) to find the toroidal component of the magnetic field.

 We note that Eq. (\ref{eq:ind_dip}) has been solved in the context 
of stellar winds by Mestel (1961); his result was that the poloidal 
magnetic field and velocity field need to be parallel. Our result that 
$v_r/B_r=v_\theta/B_\theta$ (from Eq. (\ref{eq:an_pol}) with 
$\tilde{k}=0$) is in agreement with this.

\subsection{Solution method}

Before solving Eq. (\ref{eq:pde1}) we put it into a dimensionless 
form, by scaling the quantities in the following way:
 \begin{eqnarray}
r=\hat{r}\,r_{{\rm g}} \hspace{1cm} \theta=\hat{\theta} \hspace{1cm} 
\mathcal{S}=\hat{\mathcal{S}}\,\mathcal{S}_0
\end{eqnarray}
 where the hat quantities are dimensionless, $r_{{\rm g}}$ is the 
 Schwarzschild radius and $\mathcal{S}_0$ is a reference value for the 
 stream function, calculated as the value for a dipolar field on the 
 equator of the neutron star. Substituting into Eq. (\ref{eq:pde1}) 
 with $k=0$ we get: 
 \begin{equation}
\label{eq:dim1}
\frac{\mathcal{S}_0}{r_{{\rm g}}^2}\,\partial_{\hat{r}}^2 
\hat{\mathcal{S}} + \frac{\mathcal{S}_0}{r_{{\rm 
g}}^2\hat{r}^2}\,\partial_{\hat{\theta}}^2\hat{\mathcal{S}} - \left( 
\frac{\cot\hat{\theta}}{r_{{\rm g}}^2\hat{r}^2} + \frac{v_\theta}{r_{{\rm 
g}}\hat{r}\,\eta} \right) \,\mathcal{S}_0\partial_{\hat{\theta}} 
\hat{\mathcal{S}} -
\frac{v_r}{\eta}\,\frac{\mathcal{S}_0}{r_{{\rm g}}}\partial_{\hat{r}} 
\hat{\mathcal{S}} = 0
\end{equation}
\begin{equation}
\label{eq:dim2}
\partial_{\hat{r}}^2\hat{\mathcal{S}} + 
\frac{1}{\hat{r}^2}\partial_{\hat{\theta}}^2\hat{\mathcal{S}} - \left( 
\frac{\cot\hat{\theta}}{\hat{r}^2} + \frac{v_\theta\,r_{{\rm 
g}}}{\hat{r}\,\eta} \right) \,\partial_{\hat{\theta}} \hat{\mathcal{S}} - 
\frac{v_r r_{{\rm g}}}{\eta} \partial_{\hat{r}} \hat{\mathcal{S}}=0
\end{equation}
 where we go from (\ref{eq:dim1}) to (\ref{eq:dim2}) by dividing both 
sides by $\mathcal{S}_0/r_{{\rm g}}^2$. We rename the variables 
($\hat{r}=x$ and $\hat{\theta}=y$) and obtain the following dimensionless 
form for the equation:
 \begin{eqnarray}
 \label{eq:S_dim}
 \partial_{x}^2\hat{\mathcal{S}} + 
\frac{1}{x^2}\partial_{y}^2\hat{\mathcal{S}} - \left( \frac{\cot y}{x^2} 
+ \frac{v_\theta\,r_{{\rm g}}}{x\,\eta}  \right)\,\partial_y 
\hat{\mathcal{S}} - \frac{v_r\,r_{{\rm 
g}}}{\eta}\,\partial_{x}\hat{\mathcal{S}}= 0
\end{eqnarray}
 where $[v_r]=[v_\theta]=$ cm s$^{-1}$, $[r_{{\rm g}}]=$ cm, $[\eta]=$ 
 cm$^2$ s$^{-1}$, so that $v_r\, r_{{\rm g}}/\eta$ and $v_\theta\, r_{{\rm 
 g}}/\eta$ are dimensionless coefficients.

 We have solved Eq. (\ref{eq:S_dim}) with the Gauss-Seidel relaxation 
method, which uses a finite difference technique, approximating the 
operators by discretizing the functions over a grid. At any given iteration
step, the values of the stream function at the various grid points are written
in terms of values at the previous step, or at the present step in the case of
locations where it has already been updated. Details of the numerical scheme are given in the Appendix (only in the online version of the paper).

\begin{figure}[ht!]
\begin{center}$
\begin{array}{c}
\includegraphics[width=0.37\textwidth]{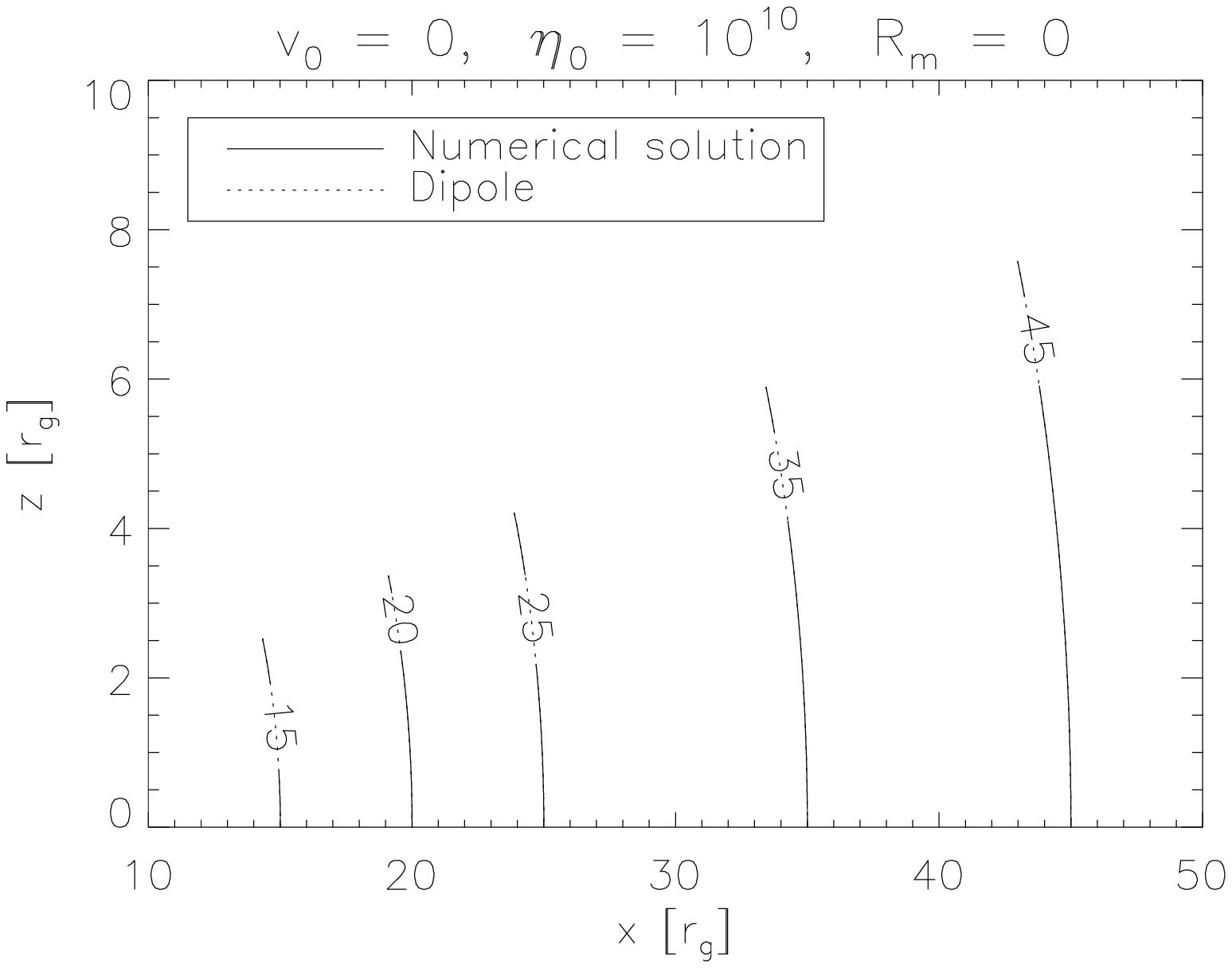}  \\ 
\includegraphics[width=0.37\textwidth]{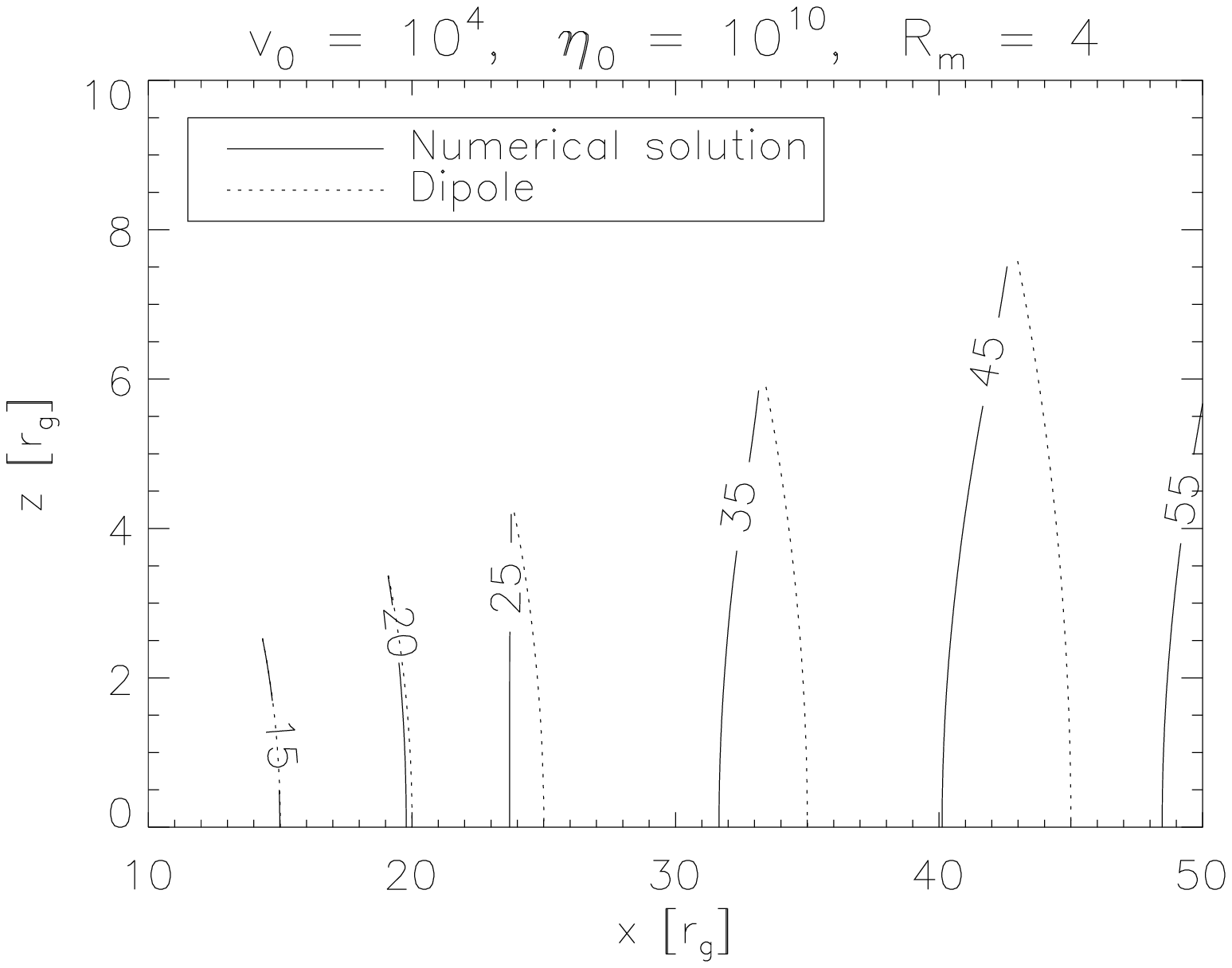}
\end{array}$
\end{center}
 \caption{Magnetic field lines from the numerical solution (solid) and 
 those for a dipole (dotted). The \emph{top panel} refers to a case with no 
 accretion, where the field remains exactly dipolar, while in the \emph{bottom 
 panel} $v_0=10^4$ cm s$^{-1}$, and the field is distorted. The diffusivity 
 $\eta_0$ has the same value in both panels ($10^{10}$ cm$^2$ s$^{-1}$). 
 If we use $\eta_0=10^{11}$ cm$^2$ s$^{-1}$ and $v_0=10^5$ cm s$^{-1}$, we 
 get the same results as shown in the bottom panel ($R_{\rm m}=4$ for 
 both).}
\label{fig:v0a}
\end{figure}

 Before applying the numerical scheme to the real problem that we want to 
solve, we performed a series of tests on the code, which are described in 
detail in the Appendix. We used several configurations, with many different
numbers of grid points, profiles for the velocity and diffusivity, initial
estimates for the stream function, locations for the outer radial boundary of
the grid and values for the iteration time step. In this way we have checked the stability and convergence, have optimized the iteration procedure and have
determined where to place the outer radial boundary of the grid (which needs to
be far enough away so that the outer boundary conditions do not significantly
influence the solution in our region of interest).

 All of the results presented in the next section have been obtained using 
a $1000\times20$ grid and with $2^{24}\sim1.7 \times 10^7$ total iterations. 
With these settings, we have always obtained residuals of the order of 
$10^{-14}$ or less, and the resolution is $\Delta x\sim0.7\,r_{\rm g}$ and 
$\Delta \theta \sim 0.5^{\circ}$.

\section{Results}\label{sec:RES}

In this section we describe how the magnetic field configuration 
changes when we modify the velocity field and the turbulent diffusivity.

\begin{figure}[ht!]
\begin{center}$
\begin{array}{c}
\includegraphics[width=0.37\textwidth]{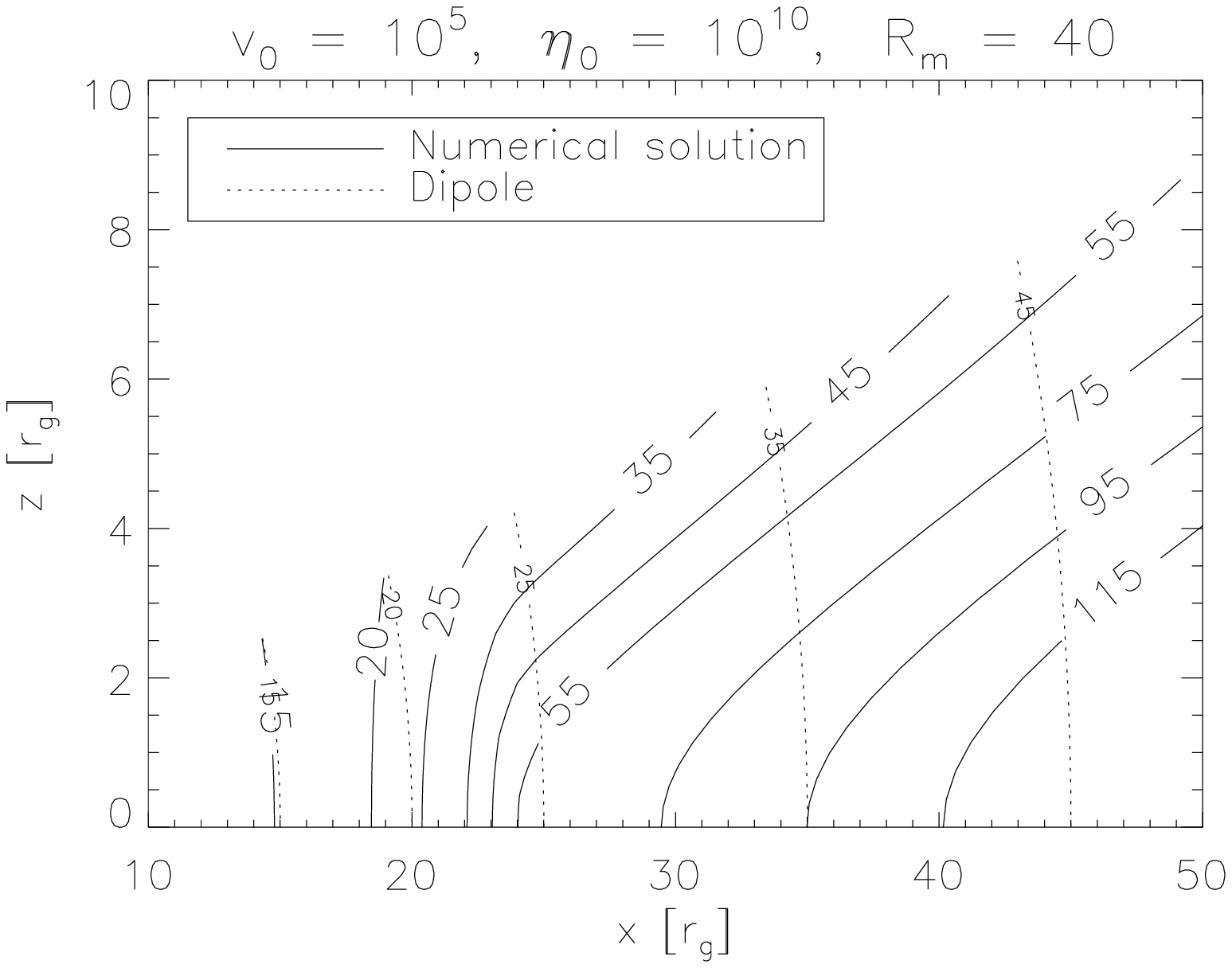}  \\ 
\includegraphics[width=0.37\textwidth]{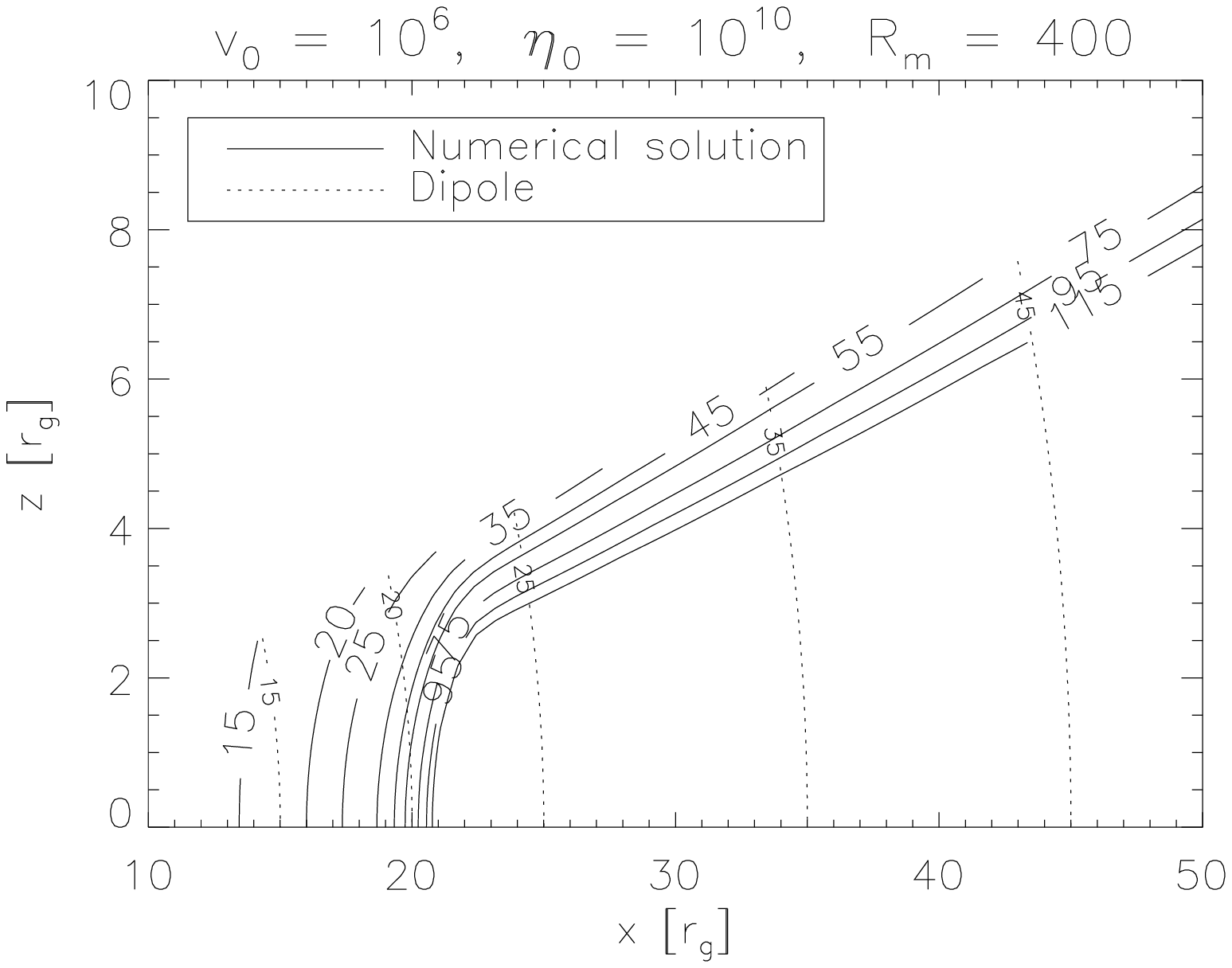}
\end{array}$
\end{center}
 \caption{The same as in Fig. \ref{fig:v0a}, but with different values 
of $v_0$. The \emph{top panel} is for a case with $v_0=10^5$ cm s$^{-1}$ while 
the \emph{bottom one} is for $v_0=10^6$ cm s$^{-1}$. If we use $\eta_0=10^{9}$ 
cm$^2$ s$^{-1}$ and $v_0=10^5$ cm s$^{-1}$, we get the same as in the 
bottom panel ($R_{\rm m}=400$ for both).}
 \label{fig:v0b}
\end{figure}

 In Figs. \ref{fig:v0a} and \ref{fig:v0b} we show the magnetic field 
lines calculated with four values of $v_0$, i.e. for different accretion 
rates\footnote{For the Shakura-Sunyaev model, the radial velocity is 
proportional to $\dot{m}^{2/5}$ if the mass of the central object and the 
viscosity $\alpha$ are kept fixed.}. For facilitating the comparison we 
show also a dipolar magnetic field (dotted). The field lines are labelled 
with the radial coordinate where the dipole field imposed at the top 
boundary would cross the equatorial plane if not distorted. We can see 
that if $v_r$ were zero, the field would not be distorted at all from the 
dipolar configuration and increasing the velocity then creates 
progressively more distortion. The degree of distortion depends on the 
location in the disc: in the inner part, where the field is strongest, it 
is most able to resist distortion; further out, the field is weaker and it 
becomes progressively more distorted. 

\begin{figure}
\begin{center}
\includegraphics[width=0.43\textwidth]{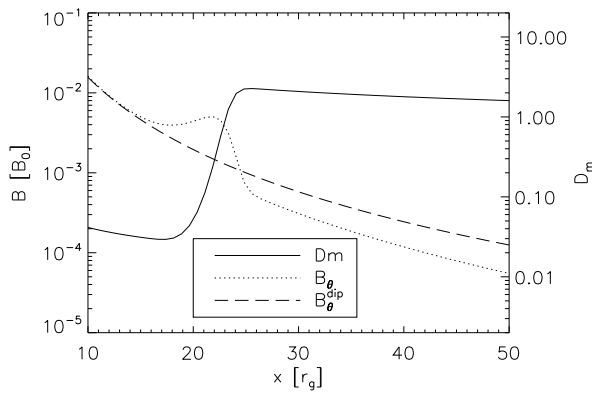}
\end{center}
 \caption{Comparison between the numerical $B_\theta$ (dotted) and 
 the dipolar one (dashed) on the equatorial plane for $R_{\rm m}=40$. The 
 solid line shows the magnetic distortion function $D_{\rm m}$. The scale 
 for this is shown on the right vertical axis, and the one for the 
 magnetic field is on the left ($B_0$ is the stellar field and is taken as 
 $3 \times 10^8$ G).}
 \label{fig:Btf}
\end{figure}

 According to the behaviour of the magnetic field lines, we can divide the 
disc into three regions: (1) an inner region, where the lines are not 
distorted very much away from the dipole; (2) an outer region, where the 
distortion can be very large and (3) the region in-between the two, which 
we call a transition region, where there is an accumulation of field lines. 
Consequently in the transition region there is an amplification of the 
magnetic field (see Fig. \ref{fig:Btf}).

 In addition to varying the radial velocity, we also considered the role of the diffusivity. The results show that when we change $\eta_0$, we get the opposite behaviour to that seen when varying the velocity, i.e. a larger $\eta_0$ gives a smaller distortion. Actually what really matters is not the velocity or the diffusivity alone but their ratio. This is not surprising since in the equation which we are solving (Eq. (\ref{eq:S_dim})) the quantities only appear in this ratio (bearing in mind that $v_\theta$ is taken to be either zero or proportional to $v_r$). In fact the magnetic Reynolds number $R_{\rm m}$, which describes the general solution of the induction equation (\ref{eq:IND}), is built from them: it is defined as $R_{\rm m}\equiv{l_0\cdot v_0}/{\eta_0}$, where $l_0$, $v_0$ and $\eta_0$ are respectively a characteristic length, velocity and diffusivity. This parameter gives the relative importance of the two terms on the right-hand side of the induction equation. For large $R_{\rm m}$, we are in the regime of ideal MHD with the magnetic field and plasma being frozen together; for low $R_{\rm m}$, instead, the field and plasma are almost decoupled and the field simply diffuses. In accretion discs the radial velocity is usually many orders of magnitude smaller than the azimuthal velocity. In our case the Reynolds numbers calculated using the two velocities differ by about five orders of magnitude, if one takes the Keplerian velocity as the characteristic $\phi$ velocity. For our present calculations however only the radial motion is relevant, since $v_\theta=0$ in the disc and is proportional to $v_r$ in the corona, while $v_\phi$ does not appear in the equation that we are solving now. The value of $R_{\rm m}$ reported in Figs. \ref{fig:v0a} and \ref{fig:v0b} is therefore the one calculated taking the characteristic velocity to be the radial velocity.

 The panels in these figures are clearly showing that the distortion of the field is proportional to the magnetic Reynolds number calculated in this way. This happens because with increasing $R_{\rm m}$ the freezing condition gets progressively stronger so that any fluid motion perpendicular to the magnetic field lines encounters more and more resistance. Therefore, since the velocity field is fixed, the magnetic field has to change. Figures \ref{fig:v0a} and \ref{fig:v0b} not only show that modifications in the magnetic field lines increase with $R_{\rm m}$, but also that their shape is consistent with what is expected from considering the flux freezing condition in the case of a conical flow (which is what we have in the disc).

 However the actual value of the magnetic Reynolds number is somewhat 
arbitrary, because in general there is no unambiguous way of defining the 
characteristic length, velocity and diffusivity of a given system. In our 
case we choose $l_0$ to be the radius of the inner edge of the disc, $v_0$ 
to be the radial velocity at the inner edge of the disc and $\eta_0$ to be 
the value of the diffusivity in the main disc region. One could also make 
a different choice for the characteristic length $l_0$, such as taking 
this to be the radius of the star or the average height of the disc; the 
trend of having larger distortions for larger values of $R_{\rm m}$ would 
of course be seen in all cases, but the switching on of the distortions 
would occur at different threshold values of $R_{\rm m}$.

We have already noted that the distortion varies with position, and so it 
is clear that a single global parameter cannot give a sufficiently 
detailed description in all parts of the system. It is therefore 
convenient to introduce a new quantity which we call the ``magnetic 
distortion function'' $D_{\rm m}$. We define this in the same way as the 
magnetic Reynolds number but, instead of taking characteristic values for 
the velocity and diffusivity, we take the local values:
 \begin{equation}
 D_{\rm m}(r,\theta)=\frac{l_0 |\mathbf{v}(r,\theta)|}{\eta(r, \theta)}
\end{equation}
 This function gives the relative importance of the two terms on the 
 right-hand side of the induction equation at every point of the disc, 
 rather than giving just a global measure as with the standard magnetic 
 Reynolds number. We then expect the advection term ($\nabla \times 
 (\mathbf{v} \times \mathbf{B})$) to dominate if $D_{\rm m}\gg1$
and the diffusion term ($\nabla \times (\eta \nabla \times 
 \mathbf{B})$) to dominate if $D_{\rm m}\ll1$. This then explains
why we find three regions inside the disc: the inner region corresponds to the 
 zone where $D_{\rm m}\ll1$, the main region to $D_{\rm m} > 1$,
and the transition region to intermediate values of $D_{\rm m}$. This 
 correspondence is made clear in Fig. \ref{fig:Btf}, where we show the 
 $\theta$ component of the magnetic field, the dipolar profile and the 
 magnetic distortion function, all on the equatorial plane. We 
 recall that the jump in $D_{\rm m}$ follows from the profile chosen for 
 $\eta$, i.e. we use a larger value of the diffusivity in the inner part 
 of the disc.

 Another important aspect of the magnetic distortion function is that 
its definition is less arbitrary than that for the standard magnetic 
Reynolds number, since it is defined using only one characteristic value, 
$l_0$. In addition there is a quite natural way for choosing $l_0$. By 
looking at Eq. (\ref{eq:S_dim}) one can see that, if we choose 
$l_0$ to be equal to our unit of length, $r_{\rm g}$, then the magnetic 
distortion function is already there in the equation (it is the 
coefficient of the partial derivative of $\mathcal{S}$ with respect to 
$x$). We can then think of $l_0$ as a quantity needed to make the ratio 
$v/\eta$ dimensionless, and the most natural choice for this is the 
characteristic scale being used as the unit length.

 Summarizing, we can describe the magnetic field configuration in the 
accretion disc by saying that magnetic field lines that enter the disc in 
the main region ($D_{\rm m} > 1$) are pushed towards the central 
object, whereas those which enter the disc in the inner region ($D_{\rm m}\ll1$) are almost unmodified. The result is that in between these two regions there is an accumulation of field lines, and so there is an amplification of the magnetic field there, as can be seen in Fig. \ref{fig:Btf}.

 In order to test the dependence of the results on the 
boundary conditions, we have experimented with several different profiles 
for the magnetic stream function outside the disc. We have used 
three additional profiles: one which gives a magnetic field with spherical 
field lines, one which gives a magnetic field with vertical field lines 
and another one which gives field lines inclined at an arbitrary angle to 
the vertical axis. We have chosen these profiles by comparison with the 
results from the simulations by Miller \& Stone (\cite{MS97}). For 
magnetic field lines entering the disc at the same locations, their shape 
within the disc varies hardly at all in the different cases (although 
the actual value of the field strength can be different). We conclude 
that the distortion of the field lines does not depend sensitively on the 
boundary conditions used and is instead mainly governed by the magnetic 
distortion function $D_{\rm m}$.

\begin{figure}
\begin{center}$
\begin{array}{c}
\includegraphics[width=0.4\textwidth]{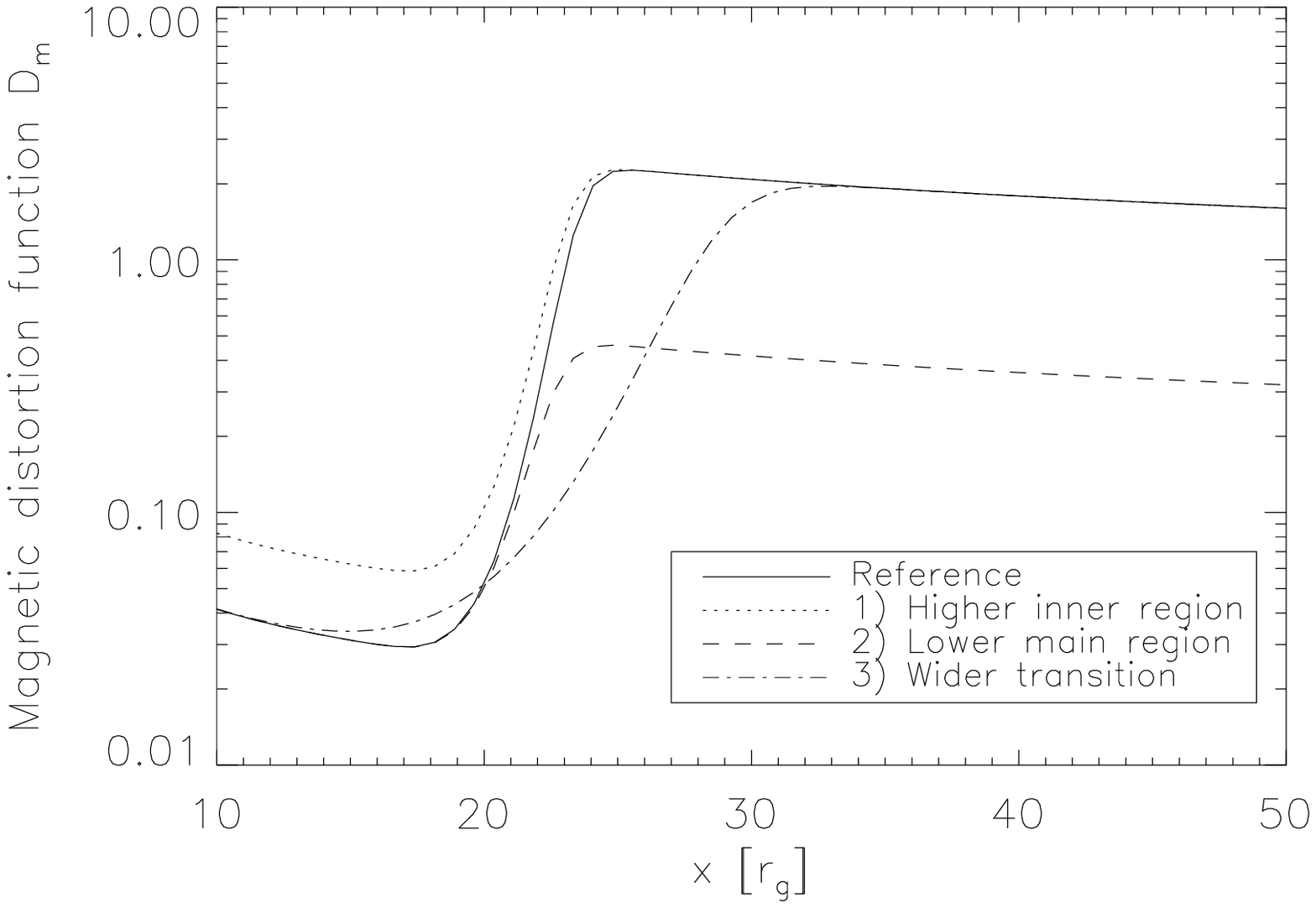}  \\ 
\includegraphics[width=0.4\textwidth]{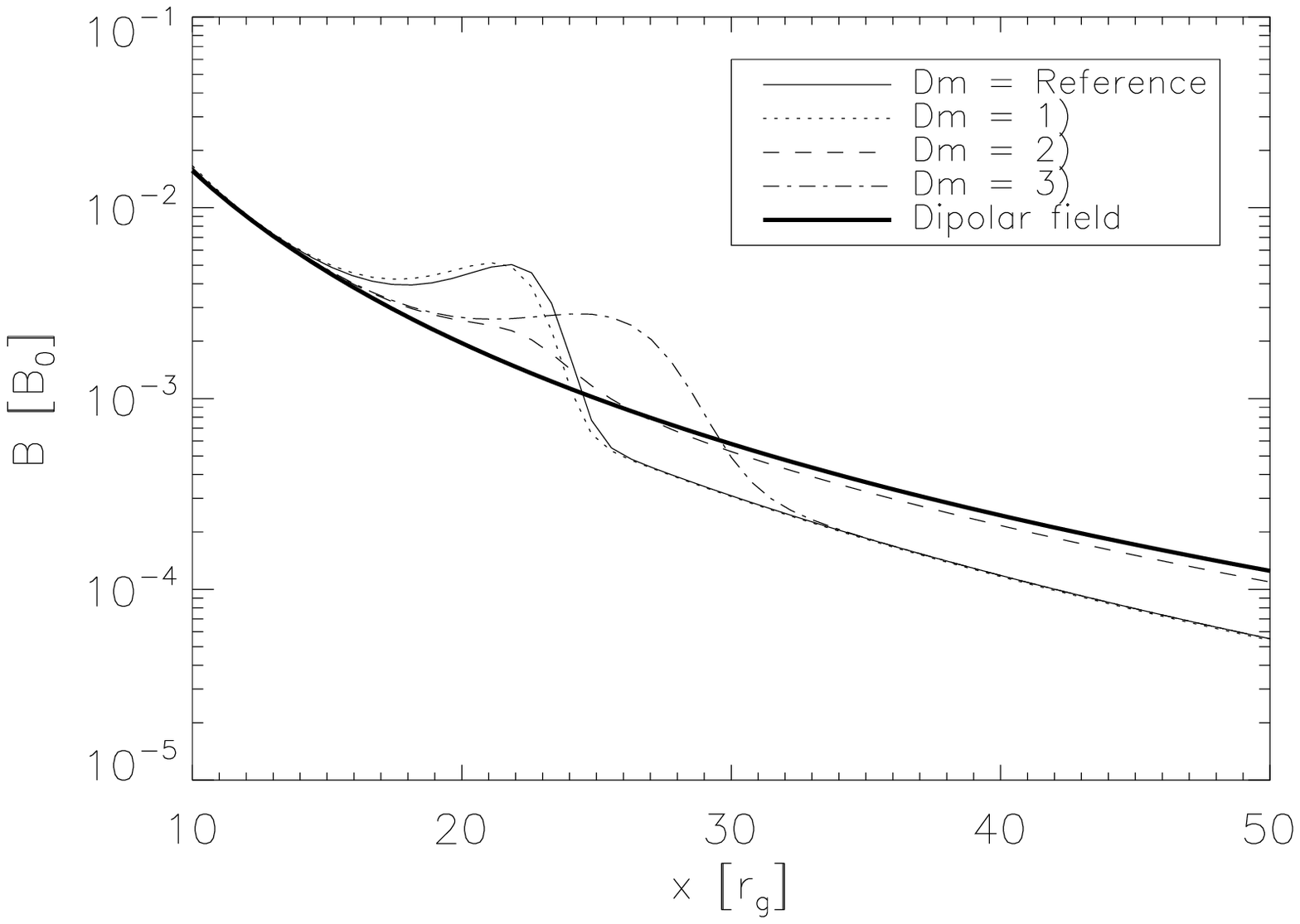}
\end{array}$
\end{center}
 \caption{\emph{Top panel}: the magnetic distortion function in the equatorial 
plane. \emph{Bottom panel}: $B_\theta$ in the equatorial plane. For both panels: the solid line shows results from the previous analysis (in the bottom 
panel the dipolar field is shown with a thick solid line); the dotted lines refer to profile 1 (larger value of $D_{\rm m}$ in the inner disc only); the dashed lines refer to profile 2 (lower value of $D_{\rm m}$ in the outer disc only); the dot-dashed lines refer to profile 3 (same values of $D_{\rm m}$ in inner and outer regions, but the transition region is wider).}
 \label{fig:Dm_Bt}
\end{figure}

 In order to better understand the influence of the magnetic distortion 
function on the magnetic field structure, we varied $D_{\rm m}$ and saw 
how the field changed. We used three new profiles for $D_{\rm m}$ and 
considered the previous one as a reference. In the first profile we 
increased the value of $D_{\rm m}$ in the inner disc and left the 
rest unmodified, in the second one instead we lowered $D_{\rm m}$ in the 
outer disc and did not change the inner part, and in the last one we just 
changed the width of the transition between the low and high values of 
$D_{\rm m}$. We then calculated the poloidal magnetic field and the 
results are presented in Fig. \ref{fig:Dm_Bt}, where the top panel shows 
the different profiles of the magnetic distortion function and the bottom 
one shows the $\theta$ component of the magnetic field, referring to the 
equatorial plane in both cases.

\begin{figure}
\begin{center}
\includegraphics[width=0.4\textwidth]{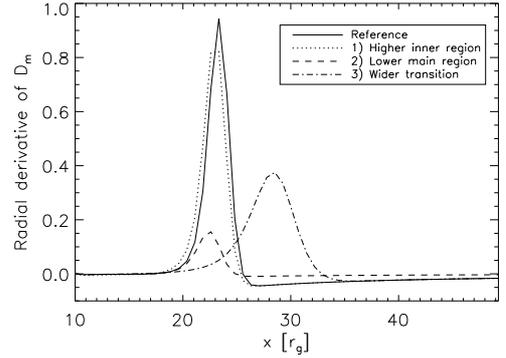}
\end{center}
 \caption{The radial derivative of the magnetic distortion function in the 
 equatorial plane. The profiles refer to the same case as in Fig. \ref{fig:Dm_Bt}. Comparing with the bottom panel of Fig. \ref{fig:Dm_Bt} one can see that the peaks in the magnetic field $B_\theta$ occur at almost the same locations as where $\partial_r\,D_{\rm m}$ has a maximum, and that the amplitudes of the distortions are proportional to the magnitude of the maximum of $\partial_r\,D_{\rm m}$.}

 \label{fig:dr_Dm}
\end{figure}

 Considering this figure, we can summarize the influence of the magnetic 
distortion function with four comments: (1) changing the value of $D_{\rm 
m}$ in the inner part by a factor of 5 leaves the magnetic field almost unchanged; (2) on the other hand, the magnetic field is very sensitive to the width of the transition in $D_{\rm m}$ and to its value in the outer disc; in particular (3) the position of the peak in $B_\theta$ is related to the width of the transition; and (4) the deviations away from the dipole field are mainly governed by the value of $D_{\rm m}$ in the outer disc. We can go further and consider the radial derivative of $D_{\rm m}$, which is shown in Fig. \ref{fig:dr_Dm} for all of the profiles used. From this we can see that the position of the peak of $B_\theta$ is strongly connected with the position of the maximum in $\partial_r\,D_{\rm m}$, and that the maximum amount of 
magnetic distortion is related to the height of the peak in the derivative of $D_{\rm m}$.

\section{Conclusions}\label{sec:CON}
 In this paper we have begun a systematic study of the magnetic field 
 configuration inside accretion discs around magnetised neutron stars, 
 which is intended as being complementary to the large numerical 
 simulations being carried out elsewhere. We have assumed that the star 
 itself has a dipolar magnetic field, whose axis is aligned with the 
 rotation axis, which is perpendicular to the disc plane. We have also 
 assumed that the flow is steady and has axial symmetry everywhere. Our 
 strategy was to start the analysis with a very simple model, where we 
 made the kinematic approximation, included turbulent magnetic
diffusivity and solved the induction equation numerically in full 2D, without
making any leading order expansion. This initial model will subsequently be
improved by including the magnetic back-reaction on the fluid flow.

 We have shown that it is possible to separate the calculation for the 
configuration of the poloidal magnetic field from any azimuthal quantities.
This is a key point and has the consequence that the effective magnetic Reynolds number is that calculated with the radial velocity rather than the azimuthal one (which is very much larger). We have here considered only the poloidal component; the toroidal one will be addressed in a subsequent paper.

 We have modelled the surroundings of the neutron star as being composed of
four regions (see Fig. \ref{fig:model}): the inner and the outer disc, the
corona (taken to be a layer above and below the disc) and all of the rest, which is taken to be vacuum. We suppose that the stellar magnetic field remains
dipolar until it reaches the corona. At that point it begins to feel the
presence of the fluid flow and the magnetic field lines are pushed inwards,
thus creating distortions away from the purely dipolar field.

 We have studied the response of the magnetic field to changes in the 
velocity and the diffusivity, finding distortions away from dipolar 
increasing with the radial infall velocity and decreasing with increasing 
diffusivity. The underlying behaviour is that the distortions increase 
together with the magnetic Reynolds number $R_{\rm m}$ where the ratio 
$v_0/\eta_0$ appears.

 However a single value of $R_{\rm m}$ cannot take into account any large 
changes in the magnitudes of the velocity and the diffusivity through the 
disc, since it is defined using single characteristic values. Therefore in 
order to have a sufficiently detailed description of the system, we have 
introduced a magnetic distortion function $D_{\rm m} = \frac{l_0 |\mathbf{v} (r, \theta)|} {\eta (r, \theta)}$, based on local values of the quantities concerned, so that in regions where $D_{\rm m} \gg 1$ or $D_{\rm m} \ll 1$ one should expect to have large or small distortions respectively. We expect the turbulence to be enhanced in the regions of lower density (the corona and the inner part of the disc), therefore in our model we use a larger value of $\eta$ in these zones (similarly to Rekowski et al. \cite{R00}), giving a smaller value for $D_{\rm m}$. Moreover in these regions the magnetic field should be less sensitive to the plasma flow and, given that we are in the
kinematic approximation, the only way of achieving this is to use a larger value of the diffusivity. As clearly shown in the panels of Figs. \ref{fig:v0a} and \ref{fig:v0b}, the disc can be divided into three parts: (1) an inner region, where $D_{\rm m}\ll1$ and the distortions are negligible; (2) a transition region, where $D_{\rm m}$ is rapidly increasing and magnetic field lines accumulate; (3) an outer region, in the inner part of which $D_{\rm m} > 1$ and the the distortions can be very large. Considering $D_{\rm m}$ can be very useful when analysing results of large numerical simulations.

 Comparing our results with previous literature, we can confirm the idea of
dividing the disc into two principle regions: an inner part, where the magnetic
field is strongest, and an outer part, where the magnetic field is weaker and
gently decaying. However the behaviour that we find for the field in these
regions is very different from that of the Ghosh \& Lamb model (\cite{GL79a})
and we find it convenient to include a third zone, to be considered as a
transition between the two principle ones (see Fig. \ref{fig:Btf}). In the
inner boundary layer of the GL model, the magnetic field is reduced by screening currents by a factor of $80 \%$, while in our case the field is barely modified in the first region. In the transition region instead we see a new effect: the field is amplified and has a local maximum. The properties of the maximum (i.e. its location, the peak magnitude of the field and its behaviour in the neighbourhood) are well described by the magnetic distortion function $D_{\rm m}$, in particular by its maximum value, by the width of the transition between the low and high values and by the behaviour of the radial derivative $\partial_r D_{\rm m}$ (i.e. by the location and the magnitude of its maximum). The behaviour in the outer zone is instead quite similar to that in the GL model, with the field decaying and being smaller than the dipole one at any given location.

 As regards the magnetic field geometry, our results resemble rather closely
those obtained by Miller \& Stone (\cite{MS97}) (compare our Fig. \ref{fig:v0b} with the top panels of their figure 3), despite the fact that they solved the full set of MHD equations whereas we have solved just the induction equation and with different conditions at the top of the disc. Moreover we have found that the distortion of the field lines inside the disc depends very little on the profiles outside it (i.e. on the boundary conditions). We conclude that the distortion pattern seen for the poloidal component of the magnetic field should be rather a general result, related only to the fundamental aspects of the system, and that the distortion is not at all negligible for typical values of the accretion rate (see Figs. \ref{fig:v0a} and \ref{fig:v0b}).

 In our next paper, we will calculate the toroidal component of the magnetic
field by solving Eq. (\ref{eq:Bphi}) and will study its dependence on the
angular velocity and other components of the magnetic field. Afterwards
additional elements will be included in the model, including back-reaction on
the plasma flow and the dynamo effect.

\section*{Acknowledgments}
 It is a pleasure to thank Alfio Bonanno, Claudio Cremaschini and Kostas 
 Glampedakis for stimulating discussions; this work was partly supported 
 by CompStar, a Research Networking Programme of the European Science 
 Foundation.

 \newpage
 
 \appendix
\section{The Code}\label{app:CODE}
 In this Appendix we describe the numerical code that we have used 
 to solve Eq. (\ref{eq:S_dim}) and discuss some of the tests that we 
 have performed on it.

\subsection{Description of the code}\label{app:descr_cod}

 In order to solve the 2D elliptic PDE (\ref{eq:S_dim}) we use the 
Gauss-Seidel relaxation method. If we call the elliptic operator 
$\mathcal{L}$ and the right hand side $b$, then the original equation 
becomes: $\mathcal{L}[\mathcal{S}]=b$. We turn this elliptic equation into 
a hyperbolic one by adding a pseudo time derivative; we can then consider 
the iterative procedure as a time evolution and write: 
$\partial_t\mathcal{S}=\mathcal{L}[\mathcal{S}]-b$. In our case 
$\mathcal{L}=\partial_{x}^2 + \frac{1}{x^2}\partial_{y}^2 - \left( 
\frac{\cot y}{x^2} + \frac{v_\theta\,r_{{\rm g}}}{x\,\eta} \right) 
\,\partial_y - \frac{v_r\,r_{{\rm g}}}{\eta}\,\partial_{x}$ and $b=0$.

 We approximate the operators by discretizing the functions over a grid. 
The scheme that we use for discretizing the derivatives is as follows:
 \begin{equation}
 \label{Aeq:dtheta}
 \partial_\theta \mathcal{S}|_{i,j} = \frac{\mathcal{S}_{i,j+1} -
\mathcal{S}_{i,j-1}} {2\Delta j} 
\hspace{.5 cm} 
\partial_{\theta}^2 \mathcal{S}|_{i,j} = \frac{\mathcal{S}_{i,j+1} -
2\mathcal{S}_{i,j} + \mathcal{S}_{i,j-1}} {\Delta j^2}
\end{equation}
\begin{equation}
 \label{Aeq:dr}
 \partial_r
\mathcal{S}|_{i,j} = \frac{\mathcal{S}_{i+1,j}-\mathcal{S}_{i-1,j}}{2\Delta 
i} \hspace{.5 cm} \partial_{r}^2 
\mathcal{S}|_{i,j} = \frac{\mathcal{S}_{i+1,j} - 2\mathcal{S}_{i,j} +
\mathcal{S}_{i-1,j}} {\Delta i^2}\\
\end{equation}
\begin{equation}
 \label{Aeq:dt}
 \partial_t\mathcal{S}=\frac{\mathcal{S}^{t+1}-\mathcal{S}^t}{\Delta t}
\end{equation}
 where the indices $i$ and $j$ refer to grid points along the  $x$ and $y$
coordinate directions respectively, while $t$ represents the pseudo-time or
iteration step. We use expressions (\ref{Aeq:dtheta})-(\ref{Aeq:dt}) to
discretize Eq. (\ref{eq:S_dim}) and then, isolating the term
$\mathcal{S}^{t+1}_{i,j}$, we get the iterative algorithm that we use in our
code:
 \begin{eqnarray}
\nonumber
\mathcal{S}_{i,j}^{t+1} =&& \mathcal{S}_{i,j}^{t} + \Delta t \, \left[ 
\frac{\mathcal{S}_{i+1,j}^t-2\mathcal{S}_{i,j}^t+\mathcal{S}_{i-1,j}^{t+1}}{
\Delta i^2} \right.\\
 \nonumber
&& + 
\frac{1}{x_i^2} \frac{\mathcal{S}_{i,j+1}^t - 2\mathcal{S}_{i,j}^t +
\mathcal{S}_{i,j-1}^{t+1}} {\Delta j^2} - \left( \frac{\cot y_j}{x_i^2} +
\frac{v_\theta\,r_{{\rm g}}}{x_i\,\eta}\right) \\
\nonumber
&& \times
\frac{\mathcal{S}_{i,j+1}^t - \mathcal{S}_{i,j-1}^{t+1}} {2\Delta j} \\
&& -
\left. \frac{v_r\,r_{{\rm g}}}{\eta} \frac{\mathcal{S}_{i+1,j}^t -
\mathcal{S}_{i-1,j}^{t+1}} {2\Delta i} \right]
\label{Aeq:discr_S3}
\end{eqnarray}
 We solve this proceeding from $i=1$, $j=1$; on the right hand side the terms that have already been calculated (i.e. the terms at positions $i-1$ and $j-1$) are taken at the current iteration $t+1$. Once $v_r$, $v_\theta$ and $\eta$ have been specified, a solution can then be obtained for any initial estimate of $\mathcal{S}$.

 The magnitude of the central dipole field and the accretion rate do not 
enter Eq. (\ref{Aeq:discr_S3}) directly, but they are used to 
calculate the location of the inner edge of the disc $r_{{\rm in}}$.

 For the mass and radius of the neutron star, we use the canonical values, 
$1.4 \, M_{\odot}$ and $10$ km respectively. We fix the accretion rate as 
$\dot{m} = 0.03$ (in units of $\dot{M}_{{\rm Edd}}$), giving a 
magnetospheric radius of about $10\,r_{{\rm g}}$ when $B_0 \sim 3 \times 10^8$ G, as typical for a millisecond pulsar.

\subsection{Testing of the code}
 In order to check the code for stability and convergence, to estimate 
 errors and to optimize the iteration procedure by choosing an appropriate 
 iteration step, we performed a number of tests, some of which are now 
 described.

 During this test phase we used the following values for the parameters:
\begin{itemize}
  \item magnetic field at the stellar surface: $B_0 = 3 \times 10^8$ G;
  \item size of the domain: $r_{\rm{in}} = 10 \, r_{{\rm g}}$,
$r_{\rm{out}} = 750 \, r_{{\rm g}}$, $\theta_{\rm{top}}=80^\circ$,
    \item[]$\theta_{\rm{eq}}=90^\circ$ and $\theta_{\rm{c}}=82^\circ$;
  \item radial velocity at inner edge: $v_r(r_{\rm{in}})=v_0=10^5$ cm s$^{-1}$
    \item[] which is the value obtained from Eq. (\ref{eq:vr}) when
    \item[] $(\alpha$, $\dot{m}) = (0.15$, $0.03)$ or $(0.1$, $0.07)$;
  \item diffusivity: $r_{\eta\,{\rm in}}=r_{{\rm in}}$, $r_{\eta\,{\rm
out}}=r_{{\rm out}}$, $\eta_0=10^{10}$ cm$^2$ s$^{-1}$ and
    \item[]$\eta_{\rm c}=10^{12}$ cm$^2$ s$^{-1}$;
  \item initial estimate for the magnetic stream function:
   \item[] $\mathcal{S} = r_0 \cdot \sin\theta / r^{0.5}$ (for a dipolar field
$\mathcal{S}^{\rm{dip}}=r_0\,\sin^2\theta/r$);
  \item iteration time step: $\Delta t = 3 \times 10^{-2} \Delta x \, \Delta y$.
\end{itemize}
The tests can be divided into two main groups: with and without a known 
analytic solution. For the latter, we can estimate errors by calculating 
the residuals and comparing the solutions obtained with different grid 
resolutions, while for the former we also have the difference between the 
computed values and the exact results.

\subsubsection{Test with an analytic solution}
 We consider two cases. The first has dipolar boundary conditions 
 and either no poloidal motion, or a velocity profile consistent with 
 condition (\ref{eq:vrvt}). In this case the poloidal component of the 
 field must be dipolar everywhere (we refer to this test as D, for 
 dipolar). The second case has the boundary conditions for $\mathcal{S}$ 
 set to zero. In this configuration, regardless of the profile used for 
 the velocity, $\mathcal{S}(r,\theta)=0$ is a solution in all of the 
 domain (we call this test Z, for zero). This last test allows us to
 test the code by including all of the terms that will be present when 
 solving for the cases of interest (i.e. including $v_r$, $v_\theta$ and 
 $\eta$).

 In both cases, we test two different configurations (which we call 
D1, D2, Z1 and Z2) by changing the velocity profile. In test D1 we set all 
velocities to zero; in test D2 $v_r$ is zero only in the disc while in the 
corona it is given by Eq. (\ref{eq:vr}) and $v_\theta$ is given by Eq. (\ref{eq:vrvt}). In test Z1 we consider the same velocity profile as the one that we will use for our cases of interest, given by Eqs. (\ref{eq:vr}) and (\ref{eq:vrvt}) and finally in test Z2 we use the same velocity profile as in test D2.
 
 In all of these tests we follow the same procedure: we verify the stability of the code, we estimate the error and see if it scales correctly, checking the convergence of the solution. We do this by studying how the numerical solution changes when varying the number of grid points ($N_i$ and $N_j$) and the number of iterations.

\begin{figure}
\begin{center}
\includegraphics[width=0.3\textwidth]{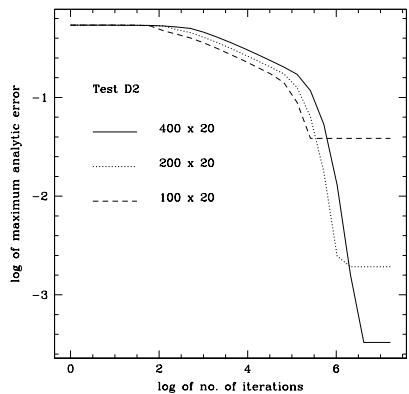}
\end{center}
 \caption{Maximum analytic error for the test D2 with three grids, which
differ only in $N_i$. Increasing $N_j$ from $20$ to $80$ produces a very small
decrease in the error, only $\sim7\%$.}
 \label{fig:test1_anerr}
\end{figure}

 We use five grids in total. When testing the dependence on $N_j$ we use: 
$200 \times 20$, $200 \times 40$ and $200 \times 80$; while when testing 
the dependence on $N_i$ we use: $100 \times 20$, $200 \times 20$ and $400 
\times 20$. For each of these grids we calculate: (i) the absolute 
difference and (ii) the relative difference, between the numerical 
solution and the analytic one at each point of the grid; and (iii) the 
root mean square (rms) of the numerical solution $\mathcal{S}$ at each 
iteration step. The results obtained are very similar for all of the five 
grids and for each of the four tests and can be summarized with the 
following four statements: (1) both the absolute error and the relative 
error have a maximum near to the inner edge $r_{{\rm in}}$ and then 
decrease quite rapidly. For the $100 \times 20$ grid, the maximum 
relative error is $\sim 15 \%$, while for the $200 \times 20$ grid it is 
$\sim 0.8 \%$ and for the $400 \times 20$ grid it is $\sim 0.15 \%$; (2) 
changing $N_j$ does not produce any visible effect: while increasing $N_j$ by a factor of 4 (from $20$ to $80$) decreases the maximum relative error only slightly ($\sim\,7 \%$), changing $N_i$ from $100$ to $400$ has a much greater effect, giving a decrease in the error of two orders of magnitude; (3) the reduction in the rms and in the maximum error becomes progressively smaller with increasing $N_i$, thus showing that we have convergence of the numerical solution; (4) using a sparser grid gives smaller errors at the beginning and during the relaxation process, however if one keeps iterating until the 
saturation level is reached, then the error with sparser grids is larger 
than with denser grids (suggesting that this problem could be suited for a 
multigrid approach). Regarding statements (2), (3) and (4), see figure 
\ref{fig:test1_anerr}.

\subsubsection{Test with an unknown solution}\label{sec:unk_sol}
\begin{figure}
\begin{center}
\includegraphics[width=0.3\textwidth]{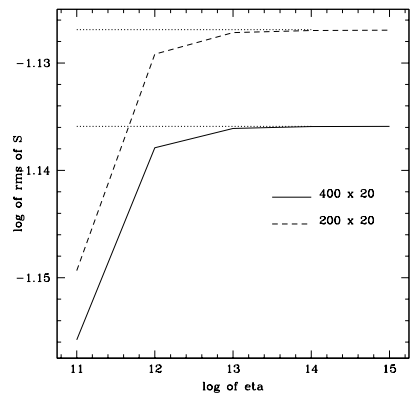}
\end{center}
 \caption{The rms of the numerical solution for different values of $\eta$ 
 after $2^{24}$ iterations. When $\eta\rightarrow\infty$, the rms should 
 reach the value for a dipole, which is plotted with a dotted line for 
 both grids.}

 \label{fig:test3_eta_rms}
\end{figure}

 Here we use the same configuration as the one that we will use for our cases of interest, i.e. dipolar boundary conditions and velocity field given by Eqs. (\ref{eq:vr}) and (\ref{eq:vrvt}). Even if we do not know the solution for this setup, we know from Eq. (\ref{eq:pde1}) that it has to approach a dipolar field when the coefficients $\frac{v_\theta}{r\,\eta}$ and $\frac{v_r}{r\,\eta}$ both go to zero. In order to test this we have considered five configurations, each with a different value of $\eta_0$ ranging between $10^{11}$ and $10^{15}$ cm$^2$ s$^{-1}$. Fig. \ref{fig:test3_eta_rms} shows clearly that for increasing $\eta$, the rms of the numerical solution is approaching that for a dipole calculated on the same grid.

 For these five configurations we also performed the tests previously described, i.e. the ones regarding changing the grid and comparing the errors and the rms. The results are again similar and confirm the four statements made earlier.

\subsubsection{Other tests}\label{sec:other_tests}
 We used the configuration of test D2 for checking three further
 aspects: (1) determining the importance of the initial estimate for 
 $\mathcal{S}$, investigating which values to choose; (2) for the location 
 of the outer radial boundary; and (3) for the iteration step.

 The kind of algorithm which we are using to solve Eq. (\ref{eq:pde1}) needs an initial estimate for the solution. According to how good or bad this estimate is with respect to the correct solution, one needs a smaller or larger number of iterations for completing the process. In order to show this and also to demonstrate that the final solution does not depend on the initial profile, we used four initial estimates for the magnetic stream function $\mathcal{S}$: (1) a constant value; (2) a Gaussian profile (centred on $r=100\,r_{{\rm g}}$ and with a width of $20\,r_{{\rm g}}$); (3) a profile increasing with $r^3$ (this gives $B_r$ and $B_\theta$ increasing linearly with $r$); and (4) the profile which gives a dipolar magnetic field. As expected, in all of the cases the final solution is the same, even for configuration (3), and the number of iterations required to reach saturation changes and goes from $0$ for case (4) to $2^{20}$ for cases (1) and (2) and to $2^{26}$ for case (3).

 As mentioned in Sects. \ref{sec:MOD} and \ref{app:descr_cod}, our 
region of physical interest goes from the inner edge of the disc $r_{{\rm 
in}}$ out to the light cylinder $r_{{\rm lc}}$. Since we do not want the 
solution in this region to be influenced by the outer boundary condition, 
we ran some tests using different values for the radius of the outer edge 
$r_{{\rm out}}$ and then compared the numerical solutions in the region of 
interest. We used the same setup as for the real problem that we were 
wanting to solve and six values of $r_{\rm{out}}$ ($150\,r_{{\rm g}}$, 
$200\,r_{{\rm g}}$, $250\,r_{{\rm g}}$, $300\,r_{{\rm g}}$, $500\,r_{{\rm 
g}}$ and $750\,r_{{\rm g}}$). We found that the difference within the 
region of interest between the numerical solutions obtained using two 
subsequent values of $r_{\rm{out}}$ became progressively smaller, until 
one could barely distinguish the different solutions. We decided to put 
the outer boundary at $750\,r_{{\rm g}}$ for the physical analysis; this 
gives results differing from those with $r_{\rm{out}}=500\,r_{{\rm g}}$ by 
less than about $5 \%$.

 Finally we considered varying the iteration step size, i.e. the $\Delta 
t$ in Eq. (\ref{Aeq:discr_S3}), that is written as $c \, \Delta r \, \Delta \theta$. There is no simple argument of principle that can be used to determine the best value for $c$, therefore we determined it experimentally. We considered the same configuration as in test D2 and ran it several times varying only the value of $c$, going from $0.025$ upwards. We found that the final asymptotic error was always the same, but that the number of iterations required to relax to the final solution was changing, decreasing as $c$ increased. However there is an upper limit: when $c>c_{\rm{max}}=1.25$ the numerical solution diverges. Transferring this condition to $\Delta t$, we obtain $\Delta t_{\rm \mbox{{\tiny Max}}} = 8.5 \times 10^{-3}$. We can then change the way in which the iteration step size is calculated in the code and write: $\Delta t = n \Delta t_{\rm \mbox{{\tiny Max}}}$, with $n$ always smaller than $1$. We find that using the value $n=0.95$ is a good compromise in minimizing the 
number of iterations and preserving the code stability.

\label{lastpage}


\begin{thebibliography}{99}

  \bibitem[2000]{AP00}  Agapitou V. \& Papaloizou J.C.B. 2000, MNRAS, 317, 273
  \bibitem[1987]{C87}   Campbell C.G. 1987, MNRAS, 229, 405
  \bibitem[1992]{C92}   Campbell C.G. 1992, GApFD, 63, 179
  \bibitem[1977]{GL77}  Ghosh P., Lamb F.K. \& Pethick C.J. 1977, ApJ, 217, 578
  \bibitem[1979a]{GL79a} Ghosh P. \& Lamb F.K. 1979a, ApJ, 232, 259
  \bibitem[1979b]{GL79b} Ghosh P. \& Lamb F.K. 1979b, ApJ, 234, 296
  \bibitem[2007]{KR07}  Kluzniak W. \& Rappaport S. 2007, ApJ, 671, 1990
  \bibitem[2008]{KR08}  Kulkarni A. K. \& Romanova M. M. 2008, MNRAS, 386, 673
  \bibitem[1961]{M61}   Mestel L. 1961, MNRAS, 122, 473
  \bibitem[1997]{MS97}  Miller K. A. \& Stone J. M. 1997, 489, 890
  \bibitem[2002]{Rom02} Romanova M. M., Ustyugova G. V., Koldoba A. V. \&
Lovelace R. V. E. 2002, ApJ, 578, 420
  \bibitem[2000]{R00} Rekowski M.v., Ruediger G. and Elstner D. 2000, A\&A,
353, 813
  \bibitem[1973]{SS73}  Shakura N.I. \& Sunyaev R.A. 1973, A\&A, 24, 337
  \bibitem[1987]{W87}   Wang Y.-M. 1987, A\&A, 183, 257

\end{thebibliography}
\end{document}